\documentclass{cernyrep}
\begin{document}
\title{A model for multifragmentation in heavy-ion reactions}
\author{A. Ferrari$^1$, M.V. Garzelli$^{2,3 }$\thanks
                 {Corresponding author e-mail: {\texttt{Maria.Garzelli@mi.infn.it}} $\,\,\,\,$ 
Proceedings of a talk 
     presented at the $12^{\mathrm{th}}$ International Conference on Nuclear 
     Reaction Mechanisms, Varenna, Italy, June 15 - 19 2009}, 
        P.R. Sala$^2$}
\institute{$^1$CERN, Geneva, Switzerland \\
           $^2$INFN \& Universit\`{a} di Milano, Dipartimento di Fisica, 
           Milano, Italy \\
           $^3$Universidad de Granada, Departamento de 
                F\'{i}sica Te\'orica y del Cosmos, Granada, Spain}
\maketitle

\begin{abstract}
From an experimental point of view, clear signatures of
multifragmentation have been detected by different experiments.
On the other hand, from a theoretical point of view, many different models,
built on the basis of totally different and often even contrasting
assumptions, have been provided to explain them.
In this contribution we show the capabilities and the shortcomings
of one of this mo\-dels, a QMD code developed by us
and coupled to the nuclear
de-excitation module taken from the multipurpose transport and
interaction code FLUKA, in re\-pro\-du\-cing 
the multifragmentation
observations recently reported by the INDRA collaboration for
the reaction Nb + Mg at a 30 MeV/A projectile bombarding energy.
As far as fragment production is concerned, 
we also briefly discuss the isoscaling technique
by considering reactions characterized by a different
isospin asymmetry, and we explain how the 
QMD + FLUKA model can be applied to obtain information on the slope 
of isotopic yield ratios, which is crucially related to the
symmetry energy of asymmetric nuclear matter. 
\end{abstract}

\section{Features of multifragmentation}
\label{section1}

When studying heavy-ion collisions at non-relativistic energies,
multifragmentation can be observed for the most central ones,
in a range of projectile-ion bombarding energies from tens MeV/A up
to a few hundreds MeV/A, depending of the properties of the
nuclei under consideration.
Many issues of this phenomenon are still under discussion, in
particular con\-cer\-ning the stage at which it occurs
in the evolution of a reaction, e.g. if a nuclear
system undergoing multifragmentation is or not equilibrated,
how a simultaneous break-up in multiple fragments can occur,
and if the multifragmentation is the result of a phase-transition.

According to the currently most believable scenario, 
during the overlapping stage of heavy-ion collisions (typical time $\simeq$ 
100 fm/c)
matter can undergo compression, leading to large excitation e\-ner\-gies. As
a consequence, the blob of nuclear matter
starts to expand and can go on expanding
down to sub-saturation densities ($\rho$ $\simeq$ 0.1 - 0.3 $\rho_0$, 
where $\rho_0$ 
is the normal nuclear matter density) 
and reach temperatures $\simeq$ 3 - 8 MeV, where
it becomes unstable and breaks up into multiple fragments. These
conditions are typical of a liquid-gas coexistence region~\cite{bondorf,buyuk}.

As already mentioned,
one of the open issues is if equilibration is reached in these reactions.
A statistical description of multifragmentation is based on this assumption.
A dynamical description of multifragmentation instead is not based on this
assumption.
Difficulties in coming to a non-controversial conclusion are largely 
due to the fact that most of the experimental data refer to patterns
of particles detected just in the
last stage 
of the reactions, involving channels feeded by the sequential decays which
heavily affect and modify the primary fragment distribution.

Multifragmentation can be distinguished from other decay channels on the
basis of the excitation energy: a typical scenario 
of small excitation energies (E $<$ 2 - 3 MeV/A) is characterized by the
formation of a compound-like system and by its evolution through binary 
sequential decays (evaporation/fission), whereas for high excitation 
energies (E $>$ 3 MeV/A) multifragmentation in a finite
volume and a simultaneous break-up into multiple fragments can occur.
The excitation energy is indeed related to the mass asymmetry $(A_{proj} -
A_{target})$: in case of symmetric central reactions the compression is 
responsible of the high excitation energy, 
whereas in case of asymmetric reactions only
a partial compression can occur and a large part of the excitation
energy appears in the form of thermal energy~\cite{singh}.
In all cases, multifragmentation is tipically driven by the following
expansion. 

Due to its mentioned features, multifragmentation, occurring during the
phase of expansion of the nuclear system formed by an ion-ion (central)
collision, allows to study the nuclear Equation of State (EoS) 
at subnormal nuclear densities. In particular,
it is possible to infer useful information concerning the symmetry energy
and its density dependence, by investigating the isotopic yield distributions 
of the emitted fragments. The isoscaling technique, based on the analysis
of isotopic yield ratios obtained in reactions with a different isospin
asimmetry, has been developed with this purpose.

After an overview of the models to study 
multifragmentation in Section~\ref{section2},
and a brief presentation of the one used in this work 
in Subsection~\ref{subsection2.1}, examples of
its application in the isoscaling technique and in the reconstruction
of multifragmenting sources at energies of a few tens MeV/A
are provided in Subsection~\ref{subsection3.1} 
and~\ref{subsection3.2}, respectively. Finally, our 
perspectives on further applications of our model are drawn in 
Section~\ref{section4}.

\section{Models to study multifragmentation}
\label{section2}

One can distinguish between 
\begin{itemize}
\item  Dynamical Models: some of them are 1-body approaches, inspired to
the BUU/BNV/Landau-Vlasov transport theory.
Alternatively, n-body approaches have been developed, such as the 
QMD/AMD/FMD. n-body approaches are very powerful in the description of 
the simultaneous break-up of a nuclear system in multiple fragments, 
since they preserve correlations among nucleons.

\item Statistical Models: 
they assume to work with an
equilibrated excited source at freeze-out (thermal equilibrium).
Taking into account that 
the nuclear system undergoes an expansion, leading to decreasing
densities, down to subnormal values, the freeze-out~\cite{trautmann} 
occurs when the mutual nuclear interaction among
fragments can be neglected. 
Statistical models have been worked out both in the 
grand-canonical framework (see e.g. Ref.~\cite{chauduri}) 
and in the micro-canonical framework. The most widespread among the last ones 
is the SMM~\cite{botvina,bondorf,gupta} 
and its modifications ISMM~\cite{tan} and SMM-TF~\cite{souza}.
\end{itemize}

We emphasize that the onset of
multifragmentation according to dynamical models is different from 
the description of multifragmentation according to statistical models.
In fact, in the statistical models a source in thermal equilibrium is 
assumed to fragment.
This means that memory effects concerning how the source has been
originated are neglected. On the other hand, in the dynamical models 
multifragmentation is a fast process: the involved
nucleons have not the time to come to equilibrium.
Fragments originate from the density fluctuations (nucleon-nucleon
correlations) due to collisions in the ion-ion overlapping stage, 
which survive the expansion phase (memory effects).
The chemical composition of hot fragments is expected to play a role 
in helping to disentangle the nature (dynamical / statistical) of the
multifragmentation mechanism~\cite{milazzo}.

Different models reproduce different features of the collisions with different
success. A mixed model, inspired to the QMD dynamical approach to describe
the fast stage of ion-ion collisions and to a statistical approach to describe
the further decay of the multiple primary excited fragments produced by QMD 
down to their ground state, has been used to obtain the results 
presented in this work. 
Due to the crucial role of dynamics, as supported by our results,   
in the following we mainly concentrate on the description 
of the dynamical aspects of multifragmentation.

\subsection{QMD/AMD approaches}
\label{subsection2.1}

In these microscopic models a nucleus is considered
a set of mutually interacting nucleons. The propagation of each nucleon occurs
according to a classical Hamiltionian with quantum effects~\cite{aiche}.
In particular, nucleons are described by gaussian wave packets.
Each of them moves under the effects of a potential given by the sum of
the contribution of all other nucleons (2-body effects).
Furthermore, when two nucleons come very close to each other, they can
undergo elastic collisions (nucleon-nucleon stochastic 
scattering cross-sections) with Pauli blocking.

A proper treatment of antisymmetrization is implemented in AMD~\cite{ono}. 
On the other hand, QMDs do not provide any antisymmetrization of the nuclear
wave-function. An approximate effect can be obtained through the inclusion
in the Hamiltonian of a Pauli potential term, or through the implementation 
of specific constraints.

Still open questions in molecular dynamics approaches concern the
functional form of the nucleon-nucleon potential (each working group
who developed a molecular dynamics code has its preferred choice of terms),
the potential parameters and their relation to the nuclear matter EoS.
Nowaday, many groups prefer parameter sets leading to a soft EoS.
Anyway, there are open questions concerning the symmetry term~\cite{li,baran}. 
In particular,
a stiff dependence for this term means that the symmetry energy always
increases with increasing densities. On the other hand, a soft dependence 
means that the symmetry energy decreases at high densities. At present, a stiff
dependence seems more reliable than a soft one.
Many uncertainties come from the fact that our observations are mainly based on
symmetric nuclear matter (N / Z $\simeq$ 1) near normal nuclear density 
($\simeq$ 0.16 $\mathrm{fm}^{-3}$), since it is difficult to obtain
highly asymmetric nuclear matter in terrestrial laboratories. 
On the other hand 
these studies are crucial to understand features of
astrophysical objects (such as neutron star formation and structure),
where conditions of extreme neutron-proton asymmetry can be present.

Other open issues concern the gaussian width, the use of 
in-medium nucleon-nucleon cross-sections instead of free nucleon-nucleon
cross-sections (in QMD the free choice is usually implemented, whereas in 
the AMD the in-medium choice has been implemented), the question
of how long the dynamical 
simulation has to be carried over and the problem of the development
of a fully relativistic approach (on the last point see e.g. 
Ref.~\cite{mancusi}).

A QMD code has been developed by us~\cite{mvg} in fortran 90. 
It includes a 3-body repulsive potential and a surface term (attractive at
long distances and repulsive at short distances).
Pauli blocking is implemented by means of the CoMD constraint~\cite{papa}. 
Neutron and proton are fully distinguished by means of a simmetry term and
an isospin dependent nucleon-nucleon stochastic scattering cross-section.
The kinematics is relativistic and attention is paid to the conservation
of key quantities (total energy/momentum, etc.) in each ion-ion collision.
Simulations are performed by means of our code from the ion-ion overlapping 
stage up to t $\simeq$ 200 - 300 fm/c (fast stage of the reaction). 
The description of the de-excitation of the excited fragments present 
at the end of the fast stage is obtained through the 
coupling of our QMD with the statistical model taken from the PEANUT module 
available in the FLUKA Monte Carlo code~\cite{fluka0,fluka1,fluka2,fluka3} 
in a version for the g95 compiler.
Up to now, the QMD + FLUKA interface has been tested in the 
collisions of ions with charge up to Z=86 (radon isotopes), providing
interesting results (see e.g. Ref.~\cite{nd2007} and references therein).

\section{Results}

\subsection{Isospin dependence in fragment production: application of the
isoscaling technique}
\label{subsection3.1}

The isoscaling technique, already mentioned in Section~\ref{section1},
is based on ratio of yields taken in multifragmentation reactions with
similar total size, but different isospin asymmetries 
($N - Z$) / ($N + Z$)~\cite{tsang, ono2}:
\begin{eqnarray}
 R_{21}(N,Z) = Y_2(N,Z) / Y_1(N,Z) = Const \,\,  \exp( A_{coeff} N + B_{coeff} Z ) \, .
\label{eq1}
\end{eqnarray}
The numerator of this formula 
refers to the yield of a given fragment ($N,Z$) obtained
from a neutron rich nucleus-nucleus reaction system,  
whereas the denominator refers to the yield of the same fragment 
from a neutron poor (more symmetric) reaction at the same energy.
$A_{coeff}$ is related to the symmetry energy and is increasingly 
larger for couple of
reactions with increasingly different isospin composition $N/Z$.

In particular, we have considered the neutron rich systems 
Ar + Fe ($N/Z$ = 1.18) and Ar + Ni ($N/Z$ = 1.13) with respect to the neutron
poor system Ca + Ni ($N/Z$ = 1.04).
Among other authors, these systems have been previously studied 
by~\cite{shetty} 
(see also Ref.~\cite{shetty2,wuenschel}).

Isotopic yield ratios for light fragment (Z $\le$ 8) emission 
have been obtained from our QMD + FLUKA simulations for the couple of reactions
Ar + Ni / Ca + Ni and Ar + Fe / Ca + Ni at 45 MeV/A projectile 
bombarding energy.
When plotted in the logarithmic plane, isotopic yield ratios for each
fixed $Z$ turn out to be approximately linear, with a slope given 
by a $A_{coeff}$, as expected from Eq.~(\ref{eq1}).
As for the isoscaling parameter $A_{coeff}$, our simulations
give the following insights:
\begin{itemize} 
\item The results of our analysis are quite sensitive to the number of isotopes
 included in the linear fit, at fixed $Z$ (i.e. to the goodness of the gaussian
 approximation to the fragment isotopic distribution).
\item   
 $A_{coeff}$ differs with $Z$, in agreement with~\cite{tsang}, 
 which claims that isoscaling is observed for a variety of reaction
 mechanisms, from multifragmentation to evaporation to deep inelastic 
 scattering, with different slopes in the logarithmic plane.
\item
 $A_{coeff}$ is larger for the couple of reactions with larger difference
 in the isospin compositions ($N_1/Z_1$ - $N_2/Z_2$).
\item 
  Our average values $A_{coeff}$ = 0.18 for Ar + Ni / Ca + Ni 
  and $A_{coeff}$ = 0.31 for Ar + Fe / Ca + Ni are larger than the experimental 
  values~\cite{shetty}, but the comparison is not so meaningful, since it is 
  largely affected by the fact that we include fragments emitted 
  in all directions in our preliminary analysis, 
  whereas in the experiment only fragments 
  emitted at 44$^o$ were selected.
 \item
   $A_{coeff}$ turns out to be affected by the choice of the impact 
   parameter and decreases significantly when selecting only the most 
   central events.
 \item
   $A_{coeff, hot}$ at the end of the overlapping stage can be larger than 
   $A_{coeff}$ at the end of the full simulation by no more than 20\%, at
   least for the reaction systems under study.
\end{itemize}

As far as the emissions at preequilibrium are concerned, our simulations lead 
to the following results: 
\begin{itemize}
\item
  For central collisions of Ca + Ni the yield of emitted protons turns out
  to be larger than the yield of emitted neutrons by 20\%.  
  For central collisions of Ar + Ni and Ar + Fe, on the other hand, 
  the yield of emitted protons turns out to be lower than the yield 
  of emitted neutrons by 10 - 15\%.
\item
  For each of the three systems under study, the fragment asymmetry
  of the liquid phase $(Z/A)_{liq}$ at the end of the preequilibrium stage 
  turns out to be lower than the corresponding value at $t=0$, in qualitative
  agreement with the AMD simulations~\cite{shetty}.
\item
 No traces of isospin fractionation appear, 
 expected indeed for systems with an higher $N/Z$ content 
(e.g. $^{60}$Ca + $^{60}$Ca).
\end{itemize}

The dependence of our results on the projectile 
bombarding energy is currently under study, by considering the same 
reactions at different bombarding energies. 

\subsection{Multifragmenting source reconstruction in Nb + Mg reactions 
at 30 MeV/A}
\label{subsection3.2}

\begin{figure}[h!]
\begin{center}
\includegraphics[width=8cm]{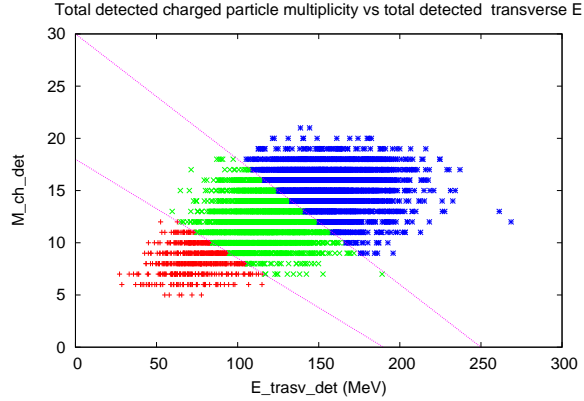}
\caption{Multifragmentation of Nb + Mg at 30 MeV/A:
           event selection and identification of different regions
           T1 (red), T2 (green) and T3 (blue) by our QMD + FLUKA simulations.
           Each point corresponds to a different ion-ion reaction 
           event in the plane identified
           by the total multiplicity of detected charged particles and
           the detected total transverse energy.}
\label{nostrafigura1}
\end{center}
\end{figure}

 Multifragmentation has been observed in Nb + Mg reactions at a 30 MeV/A
 projectile bombarding energy in an experiment performed at the INDRA 
 detector by the INDRA + CHIMERA collaborations~\cite{manduci}.
 Event selection has been performed, according to experimental cuts
 on the momentum along the beam axis, $p_{z,det} > 0.6 \, p_{z,tot}$, and on
 the angular acceptance of the INDRA detector, $4^o$ < $\theta$ < $176^o$.
 The selected events have then been assigned to different regions,
 corresponding to portions of the plane identified by
 the total transverse energy and the total 
 multiplicity of charged particles detected in each event. 
 Three regions have been
 singled out this way, as shown in Fig. 2 of Ref.~\cite{manduci}. 
 We have applied the same selection procedure by implementing
 proper cuts and filters on the simulated events obtained by
 our QMD + FLUKA.
 The selected theoretical
 events are plotted in Fig.~\ref{nostrafigura1}, which can
 be directly compared with Fig. 2 of Ref.~\cite{manduci} and turns out to
 be in good agreement.  
 The events plotted in the T1 region (red) are the less dissipative ones
 (more peripheral collisions), whereas the events in the T3 region (blue) 
 correspond to more dissipative (central) collisions. 

 For each of the three regions, average values of interesting quantities
 have been obtained both in the experiment and in the theoretical simulations.
 Our results, concerning the transverse energy, the multiplicity of
 charged particles, the velocity and the charge of the biggest residual 
 averaged over all events belonging respectively to the T1, T2 and T3
 regions are shown in Table~\ref{tabellagenerale}.
 As far as the average 
 transverse energy and multiplicity of charged particle
 are concerned, the results of our simulations turn out to be in good
 agreement with the experimental data, within the experimental uncertainties,
 in the region T1 and T2, corresponding respectively 
 to peripheral and semiperipheral collisions, whereas in case of central
 collisions the theoretical average 
 transverse energy underestimates the experimental
 one and the theoretical multiplicity of charged particles slightly 
 overestimates the experimental result. 
 On the other hand, as for the properties of the largest residual, 
 the results of the theoretical simulations show good
 agreement with experimental data especially for the most central collisions,
 belonging to the T3 region, 
 whereas for the more peripheral ones the theory overestimates the velocity
 and the charge of the largest residual.  
 These results, considered all together, seem to point out to the fact
 that in the experiment the interacting nuclei 
 are slightly more stopped than in the simulation.     

\begin{table}[h!]
\begin{center}
\caption{Multifragmentation in Nb + Mg at 30 MeV/A: 
    average values of interesting quantities 
         in the regions T1, T2 and T3.}
\label{tabellagenerale}
\begin{tabular}{p{5cm}cccc}
\hline\hline
\textbf{Region}  & \textbf{$<E_{trasv}>$} & \textbf{$<M_{tot}>$} & 
 \textbf{$<v_{maxres}>$} &
 \textbf{$<Z_{maxres}>$} \\
\hline
\multicolumn{5}{l} {Experiment (from Ref.~\cite{manduci}):}\\
\hline
 T1   &    72.9 (10.1)  &    8.1 (0.8)  &      6.6 (0.2) &        34.1 (2.3) \\
 T2   &   120.9 (10.6)  &   11.2 (0.9)   &      6.4 (0.2) &        31.9 (2.7) \\
 T3   &   176.4 (13.8)  &   13.4 (1.0)   &      6.3 (0.2) &        30.5 (2.8) \\
\hline
\multicolumn{5}{l} {Theoretical simulations (QMD + FLUKA de-exc.):} \\
\hline
T1    &    74.8      &       8.4      &       7.0        &      38.8 \\
T2    &   115.8      &      12.0      &       6.6        &      35.7 \\
T3    &   155.1      &      15.3      &       6.2        &      31.1 \\
\hline
\hline
\end{tabular}
\end{center}
\end{table}
 
\begin{figure}[ht!]
\begin{center}
\includegraphics[bb=51 51 410 200, width=15cm]{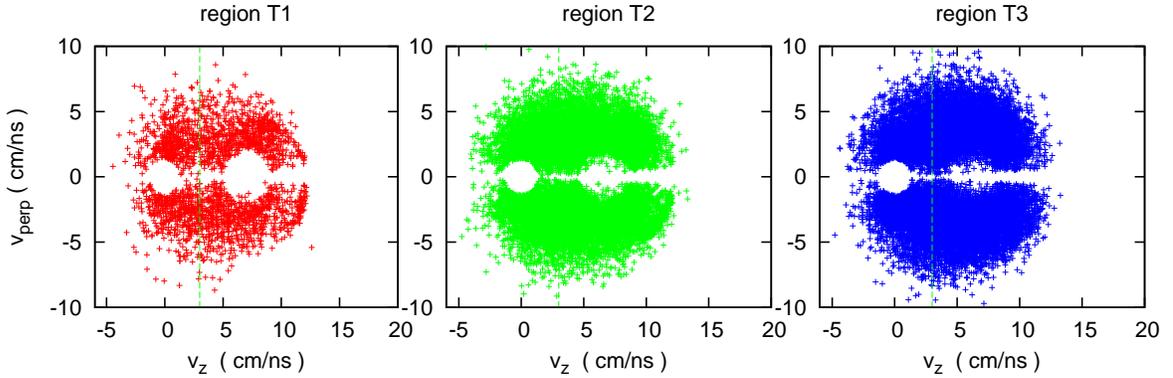}
\caption{Multifragmentation of Nb + Mg at 30 MeV/A:
           $v_{perp}$ vs. $v_{par}$ for protons emitted in the
           region T1 
(left panel), 
T2 
(central panel)
and T3 
(right panel)
, respectively, as obtained by our QMD + FLUKA simulations.
Each point in each panel correspond to a different emitted proton.
The vertical lines correspond to the velocity cuts implemented for the
reconstruction of the multifragmenting sources. 
}
\label{nostrafigura2}
\end{center}
\end{figure}

Furthermore, the considered experiment 
aim at the recostruction of the properties of the
so-called source, the blob of matter formed by compression in the
ion-ion overlapping stage, which undergoes multifragmentation.
Since the experiment detects final cold fragments, i.e. fragments
in their ground state after the de-excitation, a procedure
has been established to reconstruct the properties of the source by
using the observed properties of the final fragments and emitted protons.
In particular, the source is isolated by a selection in parallel 
velocity of different fragments (velocity cuts), by considering different
cuts in different regions. The velocity cuts implemented are
summarized in Table 2 of Ref.~\cite{manduci}.
As far as the proton are concerned, the parallel 
velocity cut is fixed to 3 cm/ns, whereas for increasingly heavier fragments
the velocity cuts are fixed to increasing values.   
The velocities of the emitted fragments are easily obtained even from 
our simulation by QMD + FLUKA, so it is possible to apply the same
procedure for the
reconstruction of the source properties even in case of our simulation. 
As an example, the velocities of the emitted protons obtained by our
simulation for events in each of the three regions 
are shown in Fig.~\ref{nostrafigura2}, by
plotting their perpendicular component $v_{perp}$
vs. their component along the beam axis $v_{par}$. 
This figure can be compared with Fig.3 of Ref.~\cite{manduci}.  
The vertical line in each panel corresponds 
to the $v_{par}$ cut implemented
in the reconstruction of the source.

Since the experiment is able to detect the charge of the emitted fragments
but not their mass, the velocity cuts can be directly used just
to obtain the charge of the source $Z_s$. 
To calculate the mass of the source $A_s$ a further hypothesis is needed.
The author of Ref.~\cite{manduci} assume that the source has the
same isotopic ratio as the projectile, i.e. $A_s/Z_s = A_{proj}/Z_{proj}$.
Source properties in the three regions T1, T2 and T3, as reconstructed 
both from the experiment and from our
simulation, are shown in Table~\ref{tabella123}. 
Since the theoretical model allows to simulate even the process of
source formation in a straightforward way, the properties of the
source can be directly obtained before its de-excitation and break-up
into multiple fragments, without using an a-posteriori reconstruction
based on velocity cuts.
If we identify the source with the biggest fragment present just at
the end of the QMD simulation, we obtain a very good agreement with the
experimental results of Ref.~\cite{manduci}, especially in the region
T1 and T2, even if the experimental results are based on the a-posteriori 
reconstruction, as can be inferred from Table~\ref{tabella123}. 
On the other hand, if we use a reconstruction procedure
analogous to the one adopted by the authors of Ref.~\cite{manduci},
we overestimate the size of the source, especially for the most central
collisions. Finally, as for the average multiplicity of IMF fragments (Z $\ge$ 3)
subsequently emitted from the source, we obtain good agreement 
with the experiment in all regions.

\begin{table}[h]
\begin{center}
\caption{Multifragmentation in Nb + Mg @ 30 MeV/A:
reconstruction of the source properties in the regions T1, T2 and T3}
\label{tabella123}
\begin{tabular}{p{5cm}ccccc}
\hline\hline
\multicolumn{6}{l} {Experiment (from Ref.~\cite{manduci}):}\\
\hline
\textbf{Region}  & \textbf{$<Z_s>$} & \textbf{$<A_s>$} & \textbf{$<M_p>$} &
 \textbf{$<M_\alpha>$} & \textbf{$<M_{frag}>$}\\
\hline
 T1  &   40.7 (2.0) & 91.2 (4.7)  &  2.0 (0.6)  &  1.1 (0.5)   &   1.2 (0.4) \\
 T2  &  42.8 (2.1) &  96.0 (4.9)  &  2.7 (0.7)  &  1.8 (0.6)   &   1.4 (0.3) \\
 T3  &   45.1 (2) & 101.3 (4.6)  &     3.1 (0.7)  & 2.5 (0.7)   &   1.6 (0.3) \\
\hline
\hline
\multicolumn{6}{l} {Theoretical simulation:}\\
\hline
\textbf{Region}  & \textbf{$<Z_s>$} & \textbf{$<A_s>$} & \textbf{$<M_p>$} &
 \textbf{$<M_\alpha>$} & \textbf{$<M_{frag}>$}\\
\hline
\multicolumn{6}{l} {Source properties reconstruction from final secondary fragments 
(QMD + FLUKA de-exc):}\\
\hline
 T1   &    42.6  &     96.2    &      2.2    &    0.4   &       1.0 \\
 T2   &   45.1   &    101.5    &     3.5     &    1.6   &       1.25 \\
 T3   &     48.5  &    109.5    &     4.2      &    3.8     &   1.5 \\
\hline
\multicolumn{6}{l} {Source properties at the end of the QMD simulation (primary fragments):}\\
\hline
 T1   &    41.0  &     91.0  & $\,$& $\,$& $\,$ \\
 T2   &   43.3  &       96.6 & $\,$& $\,$& $\,$ \\
 T3   &     47.5   &   106.1 & $\,$ & $\,$ & $\,$ \\
\hline
\hline
\end{tabular}
\end{center}
\end{table}

\section{Conclusions and perspectives}
\label{section4}

The QMD model developed in Milano and coupled to the de-excitation
module of the Monte Carlo FLUKA code has been used to study 
reactions between ions of intermediate mass which exhibit multifragmentation 
features. The results presented in this paper are encouraging, and
can be further refined by more precisely investigating up to which extent 
the statistical de-excitation process from FLUKA 
modifies the pattern of primary fragments originated dynamically by
QMD, and how the results of the simulation change when
the time of the transition from the dynamical description of the nuclear
system to a statistical description is modified. 

Further studies at non relativistic energies that
we are going to perform with our theoretical simulation tool concern:
\begin{itemize}
\item
the isospin distillation effect: it occurs in the
multifragmentation of charge-asymmetric systems, and leads to
IMF fragments (liquid) more symmetric with respect to the initial matter,
and light fragments (gas) more neutron rich. This effect is related to the 
density dependence of the symmetry energy.
\item
The bimodality in the probability distribution of the largest fragment as
a function of the mass number $A_{max}$ of the largest fragment, 
as a signature of a phase transition. Experimental data on this effect
have been obtained by the CHIMERA collaboration (see e.g. Ref.~\cite{pichon}).
\item
(Complete) fusion cross-sections (this kind of analysis has already been 
performed by other groups, e.g. by means of the ImQMD model~\cite{zhao}). 
\end{itemize}

\section*{Acknowledgements}

We wish to thank L. Manduci for enlightening comments on
the data on the reaction Nb + Mg at 30 MeV/A collected at the INDRA
detector. The QMD code developed by us and used in this study is the fruit 
of a collaboration involving many people along the years. In particular, 
we would like to mention F. Ballarini, G. Battistoni, F. Cerutti, A. Fass\`o, 
E. Gadioli, A. Ottolenghi, M. Pelliccioni, L.S. Pinsky and J. Ranft for
their support and suggestions.
The FLUKA code is under continuous development and maintainance by the FLUKA 
collaboration and is copyrighted by the INFN and CERN.


\begin{thebibliography}{99}
\bibitem{bondorf} J.P. Bondorf \emph{et al.},
\emph{Phys. Rep.} \textbf{257} (1995)  133.
\bibitem{buyuk} N. Buyukcizmeci \emph{et al.},
\emph{Phys. Rev. C} \textbf{77} (2008) 034608
[arXiv:0711.3382 [nucl-th]].
\bibitem{singh}
J. Singh \emph{et al.}, \emph{Phys. Rev. C} \textbf{62} (2000) 044617.
\bibitem{trautmann}
W. Trautmann \emph{et al.}, \emph{Phys. Rev. C} \textbf{76} (2007) 064606
[arXiv:0708.4115 [nucl-ex]]. 
\bibitem{chauduri}
G. Chauduri, S. Das Gupta, \emph{Phys. Rev. C} \textbf{76} (2007) 014619.
\bibitem{botvina}
A.S. Botvina \emph{et al.} \emph{Nucl. Phys. A} \textbf{475} (1987) 663.
\bibitem{gupta}
S. Das Gupta and A.Z. Mekjian, \emph{Phys. Rev. C} \textbf{57} (1998) 1361.
\bibitem{tan}
W.P. Tan \emph{et al.}, \emph{Phys. Rev. C} \textbf{68} (2003) 034609
[arXiv:nucl-ex/0311001].
\bibitem{souza} 
S.R. Souza \emph{et al.}, \emph{Phys. Rev. C} \textbf{79} (2009) 054602.
\bibitem{milazzo}
P.M. Milazzo \emph{et al.}, \emph{Phys. Rev. C} \textbf{62} (2000) 041602
[arXiv:nucl-ex/0002012].
\bibitem{aiche}
J. Aichelin, \emph{Phys. Rep.} \textbf{202} (1991) 233.
\bibitem{ono}
A. Ono and H. Horiuchi, \emph{Phys. Rev. C} \textbf{53} (1996) 2958
[arXiv:nucl-th/9601008].
\bibitem{li}
B.A. Li \emph{et al.}, \emph{Phys Rep.} \textbf{464} (2008) 113. 
\bibitem{baran}
V. Baran \emph{et al.}, \emph{Phys. Rep.} \textbf{410} (2005) 335
[arXiv:nucl-th/0412060].
\bibitem{mancusi}
D. Mancusi \emph{et al.}, \emph{Phys. Rev. C} \textbf{79} (2009) 014614.
\bibitem{mvg}
M.V. Garzelli \emph{et al.}, \emph{Adv. Space Res.} \textbf{40} (2007) 1350 
[arXiv:nucl-th/0611041]. 
\bibitem{papa}
M. Papa  \emph{et al.}, \emph{Phys. Rev. C} \textbf{64} (2001) 024612
[arXiv:nucl-th/0012083].
\bibitem{fluka0}
G. Battistoni \emph{et al.}, Proc. Hadronic Shower Simulation Workshop,
Batavia, Illinois, US, September 6 - 8, 2006, 
\emph{AIP Conf. Proc.} \textbf{896} (2007) 31.
\bibitem{fluka1}
A. Fass\`o \emph{et al.}, Proc. CHEP'03 Conf., La Jolla, CA, US,
March 24 - 28, 2003, (paper MOMT005) eConf C0303241 (2003), [arXiv:hep-ph/0306267].
\bibitem{fluka2}
A. Ferrari \emph{et al.}, CERN Yellow Report 2005-10, INFN/TC\_05/11, 
SLAC-R-773 (2005).
\bibitem{fluka3}
F. Ballarini \emph{et al.}, \emph{Adv. Space Res.} \textbf{40} (2007) 1339.
\bibitem{nd2007}
M.V. Garzelli, Proc. ND2007, Int. Conf. on Nuclear Data for Science and
Technology, April 22 - 27 2007, Nice, France, (editors O. Bersillon,
F. Gunsing, E. Bauge, R. Jacqmin and S. Leray, EDP Sciences) (2008), 1129
[arXiv:0704.3917 [nucl-th]].     
\bibitem{tsang}
M.B. Tsang \emph{et al.}, \emph{Phys. Rev. Lett.} \textbf{86} (2001) 5023
[arXiv:nucl-ex/0103010].
\bibitem{ono2}
A. Ono \emph{et al.}, \emph{Phys. Rev. C} \textbf{68} (2003) 051601
[arXiv:nucl-th/0305038].
\bibitem{shetty}
D.V. Shetty \emph{et al.}, \emph{Phys. Rev. C} \textbf{76} (2007) 024606
[arXiv:0704.0471 [nucl-ex]].
\bibitem{shetty2}
D.V. Shetty \emph{et al.}, \emph{Phys. Rev. C} \textbf{71} (2005) 024602. 
\bibitem{wuenschel}
S. Wuenschel \emph{et al.}, \emph{Phys. Rev. C} \textbf{79} (2009) 061602.
\bibitem{manduci}
L. Manduci \emph{et al.}, \emph{Nucl.Phys. A} \textbf{811} (2008) 93
[arXiv:0805.0975 [nucl-ex]].
\bibitem{pichon}
M. Pichon \emph{et al.}, \emph{Nucl. Phys. A} \textbf{779} (2006) 267
[arXiv:nucl-ex/0602003].
\bibitem{zhao}
K. Zhao \emph{et al.}, \emph{Phys. Rev. C} \textbf{79} (2009) 024614
[arXiv:0902.2631 [nucl-th]].
\end{thebibliography}
\end{document}